\documentclass[letter,longauth]{aa} % for the letters 
\usepackage{graphicx}
\usepackage{lscape}
\usepackage{txfonts}
\newcommand{\simnot}{\mathord{\sim}}

\begin{document}

   \title{The Gaia-ESO Survey: Discovery of a spatially extended low-mass 
   population in the Vela OB2 association\thanks{Based on observations 
   made with the ESO/VLT, at Paranal Observatory, under program 188.B-3002 
   (The Gaia-ESO Public Spectroscopic Survey)}\fnmsep\thanks{Table 1 
   is only available at CDS via anonymous ftp to cdsarc.u-strasbg.fr (130.79.128.5)
or via http://cdsweb.u-strasbg.fr/cgi-bin/qcat?J/A+A/}}

%   \subtitle{I. Overviewing the $\kappa$-mechanism}

   \author{G.~G.~Sacco\inst{\ref{inst1}}\and R.~D.~Jeffries\inst{\ref{inst2}}\and
    S.~Randich\inst{\ref{inst1}}\and E. Franciosini\inst{\ref{inst1}} \and R.~J.~Jackson\inst{\ref{inst2}}
    \and M.~Cottaar\inst{\ref{inst3}}\and L.~Spina\inst{\ref{inst1}}\and F.~Palla\inst{\ref{inst1}} \and 
    M.~Mapelli\inst{\ref{inst13}}\and E.~J.~Alfaro\inst{\ref{inst4}}\and R.~Bonito\inst{\ref{inst5},\ref{inst6}}\and F.~Damiani\inst{\ref{inst6}}\and
    A.~Frasca\inst{\ref{inst7}}\and A.~Klutsch\inst{\ref{inst7}}\and A.~Lanzafame\inst{\ref{inst8}}\and
    A.~Bayo\inst{\ref{inst9}}\and D.~Barrado\inst{\ref{inst10}}\and F.~Jim\'enez-Esteban\inst{\ref{inst10},\ref{inst12}} 
    \and G.~Gilmore\inst{\ref{inst14}}\and G.~Micela\inst{\ref{inst6}} \and A.~Vallenari\inst{\ref{inst13}}
    C.~Allende~Prieto\inst{\ref{inst18}}\and E.~Flaccomio\inst{\ref{inst6}}\and
    G.~Carraro\inst{\ref{inst15}}\and M.~T.~Costado\inst{\ref{inst4}}\and P.~Jofr\'e\inst{\ref{inst14}}\and
    C.~Lardo\inst{\ref{inst16}}\and L.~Magrini\inst{\ref{inst1}}\and L. Morbidelli\inst{\ref{inst1}}
    \and L.~Prisinzano\inst{\ref{inst5}}\and L.~Sbordone\inst{\ref{inst17}} }

   \institute{INAF-Osservatorio Astrofisico di Arcetri, Largo E. Fermi, 5, 50125, Firenze, Italy\label{inst1}
   \and Astrophysics Group, Research Institute for the Environment, Physical Sciences and Applied Mathematics, Keele University,
    Keele, Staffordshire ST5 5BG, United Kingdom\label{inst2}
   \and Institute of Astronomy, ETH Zurich, Wolfgang-Pauli-Strasse 27, 8093 Zurich, Switzerland\label{inst3}
   \and INAF-Osservatorio Astronomico di Padova, Vicolo dell'Osservatorio 5, I35122, Padova\label{inst13}
   \and Instituto de Astrof\'{i}sica de Andaluc\'{i}a-CSIC, Apdo. 3004, 18080, Granada, Spain\label{inst4}
   \and Dipartimento di Fisica e Chimica, Universit\'a di Palermo, Piazza del Parlamento 1, 90134, Palermo, Italy\label{inst5}
   \and INAF - Osservatorio Astronomico di Palermo, Piazza del Parlamento 1, 90134, Palermo, Italy\label{inst6}
   \and INAF - Osservatorio Astrofisico di Catania, via S. Sofia 78, 95123, Catania, Italy\label{inst7}
   \and Dipartimento di Fisica e Astronomia, Sezione Astrofisica, Universit\'{a} di Catania, via S. Sofia 78, 95123, Catania, Italy\label{inst8}
   \and Instituto de F\'isica y Astronom\'ia, Universidad de Valpara\'iso, Av. Gran Breta\~na 1111, Valpara\'iso, Chile\label{inst9}
   \and Depto. de Astrofísica, Centro de Astrobiología (CSIC-INTA), ESAC campus, 28691, Villanueva de la Cañada, Madrid, Spain\label{inst10}
   \and Suffolk University, Madrid Campus, C/ Valle de la Viña 3, 28003, Madrid, Spain\label{inst12}
   \and Institute of Astronomy, University of Cambridge, Madingley Road, Cambridge CB3 0HA, United Kingdom\label{inst14}
   \and Instituto de Astrof\'{\i}sica de Canarias, E-38205 La Laguna, Tenerife, Spain\label{inst18}
   \and European Southern Observatory, Alonso de Cordova 3107 Vitacura, Santiago de Chile, Chile\label{inst15}
   \and Astrophysics Research Institute, Liverpool John Moores University, 146 Brownlow Hill, Liverpool L3 5RF, United Kingdom\label{inst16}
   \and Millennium Institute of Astrophysics, Av. Vicu\~{n}a Mackenna 4860, 782-0436 Macul, Santiago, Chile\label{inst17}
   }

   \date{}

 \abstract{The nearby (distance$\sim$350-400 pc), rich Vela OB2
   association, includes $\gamma^2$ Velorum, one of the most massive binaries in the solar neighbourhood
 and an excellent laboratory for investigating the formation and early evolution of young clusters. 
 Recent Gaia-ESO survey observations have led to the discovery of two
 kinematically distinct populations in the young (10-15 Myr) cluster immediately surrounding $\gamma^2$
 Velorum. Here we analyse the results of Gaia-ESO survey observations of NGC 2547, 
 a 35 Myr cluster located two degrees south of $\gamma^2$ Velorum.
 The radial velocity distribution of lithium-rich pre-main sequence
 stars shows a secondary population that is kinematically distinct from 
 and younger than NGC 2547.
   The radial velocities, lithium absorption lines, and the positions in a colour-magnitude diagram of 
  this secondary population are consistent with those of one of the components discovered around $\gamma^2$ Velorum.
   This result shows that there is a young, low-mass stellar population spread
   over at least several square degrees
   in the Vela OB2 association. This population could have originally
   been part of
   a cluster around $\gamma^2$ Velorum that
   expanded after gas expulsion or formed in a less dense environment
   that is spread over the whole Vela OB2 region.} 
% 5 {} token are mandatory

   \keywords{Stars: formation -- Stars: pre-main sequence -- Techniques: spectroscopic -- open clusters and association: Vela OB2 -- open clusters and 
   association: NGC 2547 -- Stars: individual: $\gamma^2$ Velorum}

   \maketitle

\section{Introduction}

Observations of star forming regions (SFRs) in the solar neighbourhood
show that very young stars can be found in both high-density, gravitationally-bound 
clusters and low-density, loose stellar associations
(e.g., \citealt{Carpenter:2000, Lada:2003}). The origin of these young stellar 
populations is still being debated. All stars could form in massive dense clusters that expand 
to a larger volume after gas expulsion due to stellar winds, radiation pressure, and supernova explosions
(e.g., \citealt{Goodwin:2006, Baumgardt:2007}). Alternatively, stars may
form in a wide range of environments, 
from the dense cores of massive clusters to sparse associations \citep{Bressert:2010, Wright:2014}.
High-resolution spectroscopic surveys of SFRs and young clusters 
are powerful tools for investigating this topic, since stellar radial
velocities (RVs) allow us to characterize their current dynamical state, establishing
whether a SFR is bound or expanding into the field
(e.g., \citealt{Cottaar:2012a, Jeffries:2014}), and for identifying and disentangling
multiple populations seen in the same line of sight
(e.g., \citealt{Jeffries:2006, Alves:2012}).

The \object{Vela OB2} association consists of 93 early-type stars selected using {\it Hipparcos}
proper motions by \cite{de-Zeeuw:1999}, which are
distributed over an area of $\simnot 180~\rm deg^2$.
The mean Hipparcos distance to the Vela OB2 stars
is 410$\pm$12 pc, but its most massive star, \object{$\gamma^2$ Velorum}, is closer 
according to two interferometric observations of its orbit 
($336^{+8}_{-7}$ pc, \citealt{North:2007}; $368^{+38}_{-13}$ pc, \citealt{Millour:2007}).
The association is surrounded by a dust shell of radius $\sim$7.5 deg, centred 
on $\gamma^2$ Velorum, which is probably powering its
expansion \citep{Sahu:1992, Pettersson:2008}.
Recently, the young cluster (hereafter, the \object{Gamma Vel} cluster) around $\gamma^2$ Vel 
has been observed by the Gaia-ESO Survey \citep{Gilmore:2012, Randich:2013}. 
The results of these observations, covering about one square degree around the massive binary,
are discussed in \cite{Jeffries:2014}. They found that the cluster
is composed of two kinematically distinct populations, which are older
than previously thought ($\geq$10 Myr) and have an age difference
of $\sim$1-2 Myr. The older population (\object{Gamma Vel A}) is more concentrated around $\gamma^2$ Vel and
is roughly in virial equilibrium, while the younger population (\object{Gamma Vel
B}) is extended, clearly unbound, and supervirial; namely, the total kinetic energy of the cluster 
is higher than in a bound system in equilibrium.

The cluster \object{NGC 2547} (RA = 08h10m, DEC=-49d12m) is located about two degrees south of 
$\gamma^2$ Vel ($\sim$10 pc at the distance of $\gamma^2$ Vel). 
Several distance estimates, from 360--480 pc, 
are available in the literature \citep{Claria:1982, Shobbrook:1986, Lyra:2006, Naylor:2006,
van-Leeuwen:2009}. An accurate age of 35$\pm$3 Myr has been derived
from the lithium-depletion boundary \citep{Jeffries:2005}. 
The Gaia-ESO cluster targets are selected to well sample the age-distance-position parameter 
space. NGC2547 is particularly important given its age (one of the few clusters with an 
age between 10 and 50 Myr) and relatively short distance to the Sun. Here, we report 
the discovery of a secondary population that is younger and kinematically distinct from NGC 2547.
We show that this second component has a similar age to Gamma Vel B and is kinematically indistinguishable
from it, providing evidence for a very extended low-mass
component of the Vela OB2 
association.

\begin{figure}[ht]
\centering
\includegraphics[width=8 cm]{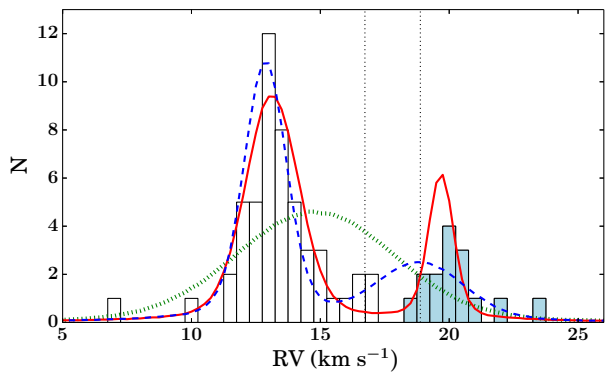}
\caption{RV distribution for all the stars with EW(Li)$>$100 m\AA.
The three lines describe the best-fit models
obtained using: (a) one stellar population with a Gaussian
distribution (green dotted line), (b) one population with a Gaussian distribution and a second one 
with the same distribution of Gamma Velorum B (blue dashed line), and (c) two populations with 
Gaussian distributions (red continuous line).
Stars with $>80$ per cent probability of belonging to population B according to 
model (b) are highlighted with blue shading. 
The vertical dotted black lines indicate the central velocity of Gamma Vel A (left) 
and Gamma Vel B (right) found by \cite{Jeffries:2014}. }
\label{fig:fig1a}
\end{figure}

\section{Observations and data}

The Gaia-ESO survey observations (see \citealt{Gilmore:2012, Randich:2013}) were carried out with the multi-object 
optical spectrograph FLAMES at the VLT, composed of a robotic 
fibre positioner feeding the GIRAFFE (R$\sim$17,000) and UVES (R$\sim$47,000) spectrographs,
with 132 and 8 fibres respectively \citep{Pasquini:2002}.

In this paper, we only use GIRAFFE data. 
The GIRAFFE candidate members are selected 
using a common strategy discussed in Bragaglia et al. (in prep.). 
Specifically, for NGC 2547 we used a two-step process: 1) we identified  
the regions including all the known cluster members in optical colour-magnitude diagrams (CMDs), 
using optical photometric data in the literature 
\citep{Naylor:2002, Jeffries:2004} and a catalogue of members
based on X-ray observations \citep{Jeffries:2006a};  2) we selected all stars 
in these regions with $11 < V < 19$ (or $10.5 < I < 16$ in the absence of $V$ photometry).
The selected catalogue of candidate cluster members includes 467 stars 
distributed over an area of about 1 deg$^2$.
We observed 450 candidates in 16 FLAMES
fields (diameter $\sim$25 arcmin each) with GIRAFFE, during two separate runs in January and 
February 2013.
Many stars were included in two or more overlapping fields and observed multiple
times.
In each field at least 20 fibres were used to acquire 
sky background spectra. All the observations were performed
with the HR15N filter covering 647--679 nm. Exposure times were 1200 or 3000 s, depending on
the brightness of the targets.

The survey spectra were processed and analysed 
by 20 working groups organized in a workflow, that is described in
\cite{Gilmore:2012}. GIRAFFE spectra were reduced
with a pipeline developed at the Cambridge Astronomical Survey Unit, which also 
calculates radial velocities (RVs) and projected rotation velocities.
A more detailed description of the data reduction and an assessment of the accuracy and the
precision of the RVs is given in \cite{Jeffries:2014} and Lewis et al. (in prep.). 
The reduced spectra of young stars 
were analysed by a dedicated working group composed of several teams,
which derive the stellar parameters (effective temperatures, gravities, 
chemical abundances, accretion tracers, as well as
equivalent widths of the Li~{\sc i}~670.8 nm line  -- EW(Li)).
The final parameters are weighted means of the results provided by the 
teams after outliers are rejected. A detailed description of the methods used
and a discussion of accuracy and precision are given by
\cite{Lanzafame:2015}. The data presented in
this paper are part of the second internal data release (GESviDR2) of July 2014
and are reported in Table 1, which is available at the CDS. 

\begin{table*}
\caption{Results from maximum likelihood modelling of the radial velocities.}
\label{tab:RV_fit}
\centering
\begin{tabular}{cccc}
\hline\hline
                               & One component & One component+$\gamma$ Vel B & Two components \\
\hline
 $RV_{A}~(\rm km~s^{-1})$      &  $14.68\pm0.44$ (14.79)  & $12.94 \pm 0.17$ (12.87)  & $13.05\pm0.21$  (13.13) \\
 $\sigma_{A}~(\rm km~s^{-1})$  &  $ 3.07\pm0.36$ (2.95)   & $ 0.83\pm 0.17$ (0.74)   & $0.97\pm0.23$   (0.95) \\
 $RV_{B}~(\rm km~s^{-1})$      &								   &   18.88                          & $18.9\pm0.95$ (19.68) \\ 
 $\sigma_{B}~(\rm km~s^{-1})$  &                                   &   1.6                            & $1.83\pm1.40$  (0.61)  \\
 $f_{A}$                       &  1.00                             &  $0.69\pm0.06$ (0.70)   & $0.72\pm0.09$  (0.75) \\
 $\ln~(L_{max})$                  &  -194.9                           &   -174.6                         & -173.81  \\
 \hline
\end{tabular}
\tablefoot{Symmetric 68 per cent confidence intervals for one parameter of interest, the value at the maximum likelihood is given in parentheses. 
The parameters $RV_A$, $RV_B$, $\sigma_A$ and $\sigma_B$ represent the mean and the standard deviation 
of the normal distributions describing the intrinsic RV distributions of the single stars in population A and B. The parameter $f_{A}$ is the       
fraction of stars belonging to population A, while $ln~L_{max}$ is the
maximum log-likelihood of the fits.}
\end{table*}

\begin{figure}[h]
\centering
\includegraphics[width=8 cm]{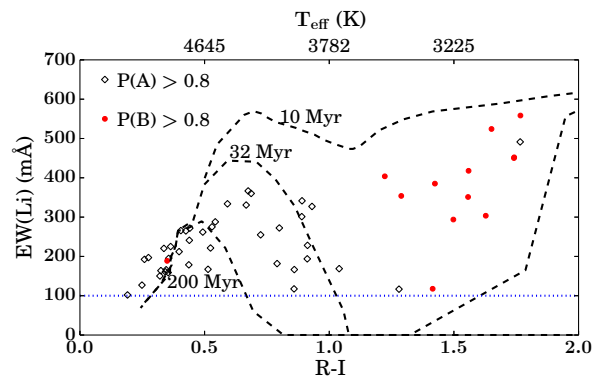}
\caption{Equivalent width of Li~670.8 nm absorption as a function of 
$R-I$ (corrected for the extinction E(B-V) = 0.06 derived by \citealt{Claria:1982}) 
for stars with $>80$ per cent probability of belonging to population A (empty diamonds)
or population B (red circles). The black dashed lines are isochrones of
Li depletion calculated from the \cite{Baraffe:1998} models
at 10, 32, and 200 Myr, and the horizontal dotted blue line indicates the threshold 
used to select the pre-main sequence stars. The x-axis on the top of the panel 
reports a scale of effective temperature derived from R-I using the \cite{Pecaut:2013} colour-temperature transformations.}
\label{fig:fig1b}
\end{figure}

\section{Population B of NGC 2547}

The strategy for the selection of targets for young clusters in the 
Gaia-ESO Survey is tailored to observe an unbiased and nearly complete sample,
but can lead to significant contamination by field stars (e.g.,
\citealt{Jeffries:2014}). To cleanly study the kinematics of the young
NGC~2547 population and search for additional young populations,
we need to remove these contaminants.

Since pre-main sequence (PMS) stars with $T_{\rm eff} \sim$4000 K completely deplete their photospheric 
lithium in $\leq$100 Myr \citep{Siess:2000, Baraffe:1998},
a filtered sample can be created by
excluding all stars with no evidence of Li absorption at 670.8 nm. 
This approach isolates a sample of young stars at cool temperatures, 
but will not select all members of NGC~2547, since
many M-type PMS stars at an age of 35 Myr will also have depleted their lithium \citep{Jeffries:2005}.

\begin{figure}
   \centering
   \includegraphics[width=8 cm]{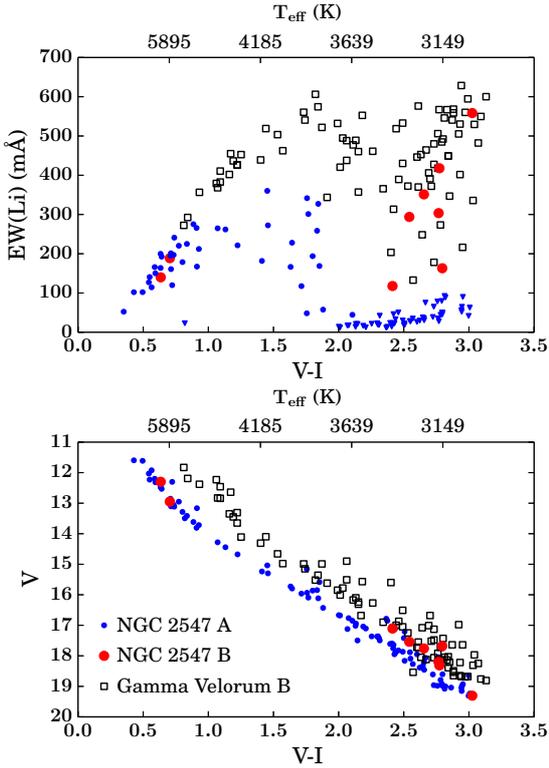}
      \caption{Comparison between the multiple populations discovered in NGC 2547 and
      Gamma Vel B. The top panel shows the EW(Li) as
      a function of $V-I$, while the bottom panel shows a $V$ vs $V-I$ CMD. Photometry has been corrected 
      according to the extinction towards NGC2547 (E(B-V)=0.06) and Gamma Velorum (E(B-V)=0.055).
      The x-axis on the top reports an effective temperature scale
      derived from V-I, using the colour-temperature transformations described in \cite{Pecaut:2013}.
      In both panels empty squares, filled red circles, and blue circles indicate the stars of Gamma Vel B, NGC 2547 B,
      and NGC 2547 A respectively. EW(Li) upper limits in NGC 2547 A are reported with downward triangles. 
      Only a subsample of stars of NGC 2547 is shown, since $V$ band photometry is not available for
      all targets (see footnote on page 4).
      }
         \label{fig:fig2}
   \end{figure}

Figure \ref{fig:fig1a} shows the RV distribution of all the observed
stars with EW(Li)$>$100 m\AA.
The distribution is characterized by two peaks at 
$\simnot 13~\rm km~s^{-1}$ and $\simnot 20~\rm km~s^{-1}$, respectively. The position of the former 
and most prominent peak is consistent with the mean RV  
of NGC 2547 ($12.8\pm1~\rm km~s^{-1}$) derived by \cite{Jeffries:2000}, while none of the 
previous spectroscopic studies of this cluster mention a kinematically distinct secondary 
population with very similar velocity to Gamma Vel B ($18.9\pm 0.5~ \rm km~s^{-1}$ from \citealt{Jeffries:2014}).
To test the hypothesis that the population associated with the second
RV peak (called NGC2547 B in
the rest of the paper) and Gamma Vel B have the same origin, we model
the RV distribution shown in Fig. \ref{fig:fig1a} using the same
maximum likelihood technique as described in detail in \cite{Jeffries:2014}, which is based on a method
developed by \cite{Cottaar:2012}. We assume that the intrinsic RV distribution
of a single population is described by a Gaussian that is broadened by binary motions
and the uncertainties in RV measurements. The RV uncertainties are calculated using equation 1 in 
\cite{Jeffries:2014}, since NGC 2547 and Gamma Vel were observed with the same
instrument and 
processed with the same pipeline. 
We tried three different models: a) a single population with mean velocity 
$RV_{A}$ and dispersion $\sigma_{A}$; b) two populations, one with mean velocity $RV_{A}$, dispersion $\sigma_{A}$,
and fraction of the total population $f_{A}$ as free parameters, and a second one with 
a fixed mean velocity ($RV_{B}= 18.88~\rm km~s^{-1}$) and dispersion ($\sigma_{B}=1.6~\rm km~s^{-1}$), 
which are the values found for Gamma Vel B by \cite{Jeffries:2014};
 c) two populations with five free parameters (mean velocities $RV_{A}$ and $RV_{B}$, dispersions
$\sigma_{A}$ and $\sigma_{B}$, and ratio between the two populations $f_{A}$).
The most likely distributions derived by the three fits are shown in 
Fig. \ref{fig:fig1a}. The best values of the free parameters and the maximum log-likelihood 
are reported in Table \ref{tab:RV_fit}. Since the models are nested, we can
perform a likelihood ratio test to establish which model describes the RV
distribution better. The test rejects model (a) when compared to the others with a probability
$>99$ per cent, while the difference in maximum log-likelihood of models (b) and (c) is not significant 
(P($\ln L_{maxA}-\ln L_{maxB})>46$ per cent). We conclude that the RV distribution of the Li-rich
members of our sample is composed of two kinematically distinct populations and that the RV distributions
of NGC 2547 B and Gamma Vel B are consistent, in agreement with the initial hypothesis.

The cluster NGC 2547 (age$\sim$35 Myr) is significantly older than Gamma Vel (age$\sim$10-15 Myr), therefore
we expect to see significant differences between the EW(Li) measured for populations A and B of NGC 2547 if the latter
is associated with Gamma Vel. In Fig. \ref{fig:fig1b}, we plot the EW(Li) as a function of 
$R-I$, together with isochrones at 10, 32, and 200 Myr derived from the \cite{Baraffe:1998} models. Different symbols
are used for stars that -- according to their RV -- are associated with 
one of the two populations with $>80$ percent confidence. As expected, in the colour range 
$1.0<R-I<1.8$, we observed very few stars belonging to NGC 2547 A, because at
an age of 35 Myr they are Li-depleted, and so were excluded from our
initial sample selection. However, we observed 
several stars belonging to NGC 2547 B in this colour range, proving that
this population is younger than 35 Myr. 
The evidence of this age difference is supported by
the presence in NGC 2547 B of a star (2MASSJ08104437-4939001)
classified as a strong accretor on the basis of its H$\alpha$ emission (EW(H$\alpha$)=78 \AA~
and a line width at 10 per cent of 440 $\rm km~s^{-1}$). Such strong accretors
are extremely rare in a 35 Myr old cluster, and there are none in NGC 2547 A.

In Fig. \ref{fig:fig2}, we compare the two NGC~2547 populations (stars
of NGC 2547 A have been selected on the basis of their RV, 
including Li-depleted stars) with Gamma Vel B\footnote{
A complete photometric catalogue in the $V$, $R$ and $I$ bands is not
available for all targets in Gamma Vel or 
NGC 2547. Therefore not all the stars plotted in Fig. \ref{fig:fig1b} are included in Fig. \ref{fig:fig2} and vice-versa.}.
The top panel shows that the EW(Li) measured in Gamma Vel B and NGC
2547 B are consistent, suggesting that they
are of similar age and younger than NGC 2547 A, because Li depletion occurs rapidly at $V-I \sim 2.5$.
The bottom panel shows that Gamma Vel B and NGC 2547 B share 
the same locus in the CMD, which, assuming that they are coeval, suggests
they are at a similar distance. The exceptions are two stars 
in the upper left-hand corner of the CMD, that match the sequence of NGC 2547 A. However, 
Li absorption cannot reliably select 
young ($<100$ Myr) stars at $T_{\rm eff}>$5000 K ($V-I<$1), so these two objects, 
which have RV ($\rm 21.65\pm0.19~and~23.10\pm0.25~ km~s^{-1}$) might be 
contaminating field stars or are possibly binary members of NGC~2547.

\section{Discussion and conclusions}

\cite{Jeffries:2014} propose
several scenarios to explain the presence of two populations around
$\gamma^2$ Velorum, 
concluding that population A is the remnant of a bound cluster formed around
the massive binary, and population B is a dispersed population from the wider Vela OB2 association.
We have discovered 15 stars that appear to belong to Gamma Vel B
and are located 2~deg 
($\sim$10 pc, assuming a distance of $\sim$340 pc) south of $\gamma^2$ Velorum,
demonstrating that Gamma Vel B extends far beyond the area 
studied by \cite{Jeffries:2014}. 
Two scenarios might explain this result: 
a) the young low-mass population observed towards NGC 2547
was part of a denser cluster around $\gamma^2$ Velorum, which -- 
after the formation of the massive 
binary and the expulsion of residual gas -- expanded into a larger volume; 
b) these stars were born in a low-density diffuse environment, which
also formed the
other members of the Vela OB2 association.

According to the relation between cluster mass and 
the mass of its most massive star proposed by \cite{Weidner:2010},
$\gamma^2$ Velorum should be surrounded by a cluster with a mass 
($\simnot 1000~ \rm M_{\sun}$), which is much higher than the present total mass ($\simnot 100~\rm M_{\sun}$) 
estimated by \cite{Jeffries:2014}. It therefore seems likely that most
of the original cluster that formed around the massive binary has
expanded into a larger volume. Assuming that the cluster started to expand
soon after the formation of the massive binary (age$\simnot 5.5\pm1$ Myr, \citealt{Eldridge:2009}) 
and considering the intrinsic velocity dispersion of Gamma Vel B ($\simnot 1.6~\rm km~s^{-1}$), 
it is possible that the fastest stars formed around $\gamma^2$ Velorum
have since moved $\sim$10 pc. This possibility has been confirmed by recent
N-body simulations aimed at investigating the origin of the two kinematically distinct populations
discovered around $\gamma^2$ Velorum \citep{Mapelli:2015}. Specifically, these simulations
show that Gamma Vel B is supervirial and extends up to a projected distance of $\sim$10 pc from the massive binary.
The predicted stellar density in the outer region of the cluster is consistent with the number  
of stars found in NGC 2547 B.
Furthermore, NGC 2547 B is only composed of low-mass stars ($R-I > 1.2$, i.e., 
below $\rm \sim0.5~M_{\sun}$).
This supports the cluster expansion hypothesis, since in this scenario
we should observe a higher concentration of stars in the cluster centre and mass segregation has been
observed in many massive clusters (e.g., \citealt{de-Grijs:2002}).

However, not all the evidence fits this first scenario. 
The presence of Li-depleted stars proves that Gamma Vel is older 
(age $>$ 10 Myr) than the massive binary. The putative expanding cluster
defined by Gamma Vel B is also offset by 2~km~s$^{-1}$ in RV from Gamma Vel A, which
appears more centrally concentrated around $\gamma^2$~Velorum. Furthermore, the other 
early-type members of the Vela OB2 association are spread over an area of radius $\sim$7-8 
degrees ($\sim$50 pc at a distance of 400 pc), 
and considering the timescale of cluster expansion, it is unlikely that 
they all formed in a much smaller region and spread out after the formation of the massive binary.
Therefore, the second scenario seems plausible, but fails to explain
why only very low-mass members of Gamma Vel B have been found towards
NGC~2547 -- the contrast in the distribution of Gamma Vel B and NGC
2547 B members in the bottom panel of Fig.~2 is striking.

%Finally, it is not clear if NGC 2547 is just a background/foreground cluster or it originated from a primary 
%episode of a long chain of sequencial star formation leading to the recent formation $\gamma^2$ Velorum.
%According to the CMD in Fig \ref{fig:fig1}, NGC 2547 is located approximately at the same distance
%of Gamma Velorum B, therefore it is possible that NGC 2547 and Gamma Velorum are part of the same star forming region.

The Gaia satellite will soon provide accurate parallaxes and proper
motions \citep{de-Bruijne:2012} for the whole
Vela OB2 region to $V \sim 19$. Therefore, it will be possible to perform an unbiased census
of the young stellar populations and study their three-dimensional spatial and kinematic structure. This 
will allow us to answer this and other questions concerning the star formation process in 
Vela OB2 uncovered by the Gaia-ESO Survey observations.

\begin{acknowledgements}
Based on data products from observations made with ESO Telescopes at the La Silla Paranal Observatory under programme ID 188.B-3002.
This work was partly supported by the European Union FP7 programme through ERC grant number 320360 and by the Leverhulme Trust through grant RPG-2012-541.
We acknowledge the support from INAF and Ministero dell'Istruzione, dell'Universit\`a e della Ricerca (MIUR) in the form of the grant "Premiale VLT 2012".
The results presented here are based on the work carried out during a visit at the University of Keele, and benefit from discussions held the Gaia-ESO workshops and 
conferences supported by the ESF (European Science Foundation) 
through the GREAT Research Network Programme. LS aknowledges the support from Project IC120009 "Millennium Institute of Astrophysics (MAS)" of Iniciativa 
Científica Milenio del Ministerio de Economía, Fomento y Turismo de Chile

\end{acknowledgements}

%-------------------------------------------------------------------

\bibliographystyle{aa}
\bibliography{/Users/sacco/LETTERATURA/bibtex_all}

\end{document}